\documentclass[twocolumn]{aastex63}
\usepackage{multirow}
\usepackage[figuresright]{rotating}
\usepackage{subfigure}
\usepackage{epic,eepic}
\usepackage{graphicx}
\usepackage{pifont}

\shorttitle{}
\shortauthors{}

\begin{document}

\title{A massive white dwarf or low-mass neutron star discovered by LAMOST}

\correspondingauthor{Song Wang}
\email{songw@bao.ac.cn}

\author{Xinlin Zhao}
\affiliation{Key Laboratory of Optical Astronomy, National Astronomical Observatories, Chinese Academy of Sciences, Beijing 100101, China}
\affiliation{School of Astronomy and Space Sciences, University of Chinese Academy of Sciences, Beijing 100049, China}

\author{Song Wang}
\affiliation{Key Laboratory of Optical Astronomy, National Astronomical Observatories, Chinese Academy of Sciences, Beijing 100101, China}
\affiliation{Institute for Frontiers in Astronomy and Astrophysics, Beijing Normal University, Beijing 102206, China}

\author{Pengfei Wang}
\affiliation{Key Laboratory of Optical Astronomy, National Astronomical Observatories, Chinese Academy of Sciences, Beijing 100101, China}

\author{Chuanjie Zheng}
\affiliation{Key Laboratory of Optical Astronomy, National Astronomical Observatories, Chinese Academy of Sciences, Beijing 100101, China}
\affiliation{School of Astronomy and Space Sciences, University of Chinese Academy of Sciences, Beijing 100049, China}

\author{Haibo Yuan}
\affiliation{Department of Astronomy, Beijing Normal University, Beijing 100875, China}
\affiliation{Institute for Frontiers in Astronomy and Astrophysics, Beijing Normal University, Beijing 102206, China}

\author{Jifeng Liu}
\affiliation{Key Laboratory of Optical Astronomy, National Astronomical Observatories, Chinese Academy of Sciences, Beijing 100101, China}
\affiliation{School of Astronomy and Space Sciences, University of Chinese Academy of Sciences, Beijing 100049, China}
\affiliation{Institute for Frontiers in Astronomy and Astrophysics, Beijing Normal University, Beijing 102206, China}
\affiliation{New Cornerstone Science Laboratory, National Astronomical Observatories, Chinese Academy of Sciences, Beijing 100101, China}

\begin{abstract}

We report the discovery of a close binary J0606+2132 (Gaia DR3 3423365496448406272) with $P_{\rm obs}=2.77$ days containing a possible massive white dwarf or a neutron star using the LAMOST spectroscopic data.
By a joint fitting of the radial velocity from LAMOST and the light curve from TESS, we derived a circular Keplerian orbit with an inclination of $i=$81.31$^{\circ}$$^{+6.26^{\circ}}_{-7.85^{\circ}}$, which is consistent with that derived from $v{\rm sin}i$.
Together with the mass of the visible star, we derived the mass of the invisible object to be 1.34$^{+0.35}_{-0.40} M_{\odot}$.
Spectral disentangling with the LAMOST medium-resolution spectra shows no absorption feature from an additional component, suggesting the presence of a compact object.
No X-ray or radio pulsed signal is detected from ROSAT and FAST archive observations.
J0606+2132 could evolve into either a Type Ia supernova or a neutron star through accretion-induced collapse if it is a white dwarf, or into an intermediate-mass X-ray binary if it is a neutron star.

\end{abstract}


\section{Introduction} 
\label{sec:intro}

Radial velocity (RV) monitoring method has been widely applied to detect compact objects.
Compared to the traditional X-ray method, the RV method can detect quiescent compact objects in binaries, which comprise the majority of binary systems including compact companions in the universe.
Recently, the RV method has found about 10 X-ray quiescent black hole (BH) candidates \citep[e.g.,][]{2014Natur.505..378C,2019Sci...366..637T,2022NatAs...6.1085S,2022A&A...664A.159M,Wang2024}, which is about half the number of BHs discovered by X-rays over the past 60 years.
A sample of star-white dwarf (WD) and star-neutron star (NS) binaries have also been discovered through the RV method \citep[e.g.,][]{2022ApJ...938...78L,2022MNRAS.517.4005M,2022NatAs...6.1203Y,2024ApJ...963..160Z,2024ApJ...964..101Z}.

For binaries including an early-type star, the large mass of the early-type star implies a massive companion, even if the system has a small binary mass function.
Thus, these binaries are promising targets for the search for massive compact objects, such as massive WDs, NSs, and BHs.
Recently, RV monitoring has identified several systems containing a massive compact object and an early-type star \citep[e.g.,][]{2014Natur.505..378C,2018ApJ...856..158K,2022ApJ...940..165Y,2022A&A...664A.159M,2022NatAs...6.1085S}.
The discovery of more compact objects can help to construct a comprehensive mass distribution, which is essential for advancing our understanding of binary evolution.

Based on data from the Large Sky Area Multi-Object Fiber Spectroscopic Telescope (LAMOST), we conducted a campaign to search for compact objects hidden in binaries including early-type stars.
This search utilized several catalogs of early-type stars \citep{2021RAA....21..288S,2022A&A...662A..66X,2022RAA....22b5009G} that were constructed using LAMOST spectra.
First, we cross-matched these catalogs with LAMOST DR9 low-resolution and medium-resolution general catalogs. 
Second, we focused on the sources which have more than two LAMOST medium-resolution spectral observations with $r$-band signal-to-noise ratio (S/N$_{r}$) greater than 5 and show clear RV variation.
Finally, we identified a close binary Gaia ID 3423365496448406272 (R.A. = 91.7051933$^o$; Decl. = 21.5425898$^o$; hereafter J0606+2132) which has a large binary mass function, suggesting the possible presence of a compact object.
This source has been recognized as a binary system in several studies \citep{2021ApJS..256...14Z,2022ApJS..258...26Z,2020ApJS..249...22T}, 
but its properties haven't been investigated in detail.
Additionally, \cite{2024ApJ...969..114L} proposed J0606+2132 as a compact object candidate based on its RV variation.
The paper is organized as follows.
In Section \ref{sec:obs_data}, we detail the spectroscopic and photometric observations.
Section \ref{sec:visible_star} presents stellar information of the visible star, including the distance, atmospheric parameters, and the mass, etc.
In Section \ref{sec:orbit}, we estimate the systematic information of J0606+2132 through RV fitting and light curve (LC) fitting.
Section \ref{sec:discussion} discusses the nature of the companion and gives possible evolution scenarios for J0606+2132.
Finally, we provide a summary in Section \ref{sec:summary}.

\section{Observations and Data Reduction} 
\label{sec:obs_data}

\subsection{LAMOST Observation} 
\label{sec:lamost_obs}

LAMOST, also named the GuoShouJing Telescope, is a reflecting Schmidt telescope, with an effective aperture of 4 m and a field of view of 5 degrees \citep{2012RAA....12.1197C,2012RAA....12..723Z}. 
It performs both low- and medium-resolution spectral observations with $\Delta \lambda / \lambda \sim$ 1800 and $\sim$ 7500, respectively. 
We obtained 29 low-resolution spectra (LRS) from 2016 January to 2022 February, covering a wavelength range from 3690 \AA \ to 9100 \AA \ \citep{2015RAA....15.1095L}, and 59 medium-resolution spectra (MRS) from 2017 November to 2021 February, with the blue and red arms covering wavelength ranges from 4950 \AA \ to 5350 \AA \ and from 6300 \AA \ to 6800 \AA, respectively \citep{2020arXiv200507210L}.
We selected the spectra with S/N$_{r}$ greater than 10, including 28 LRS and 23 MRS, and employed two methods for RV measurements.
The first way is to apply the cross-correlation function (CCF) for calculating the correlation coefficient between the template and the observed spectrum.
The RV value can be estimated using the velocity with the maximum correlation coefficient.
The wavelength range from 4000 \AA \ to 5000 \AA \ (blue band) and from 6000 \AA \ to 7000 \AA \ (red band) were utilized for LRS spectra; while the range from 4950 \AA \ to 5350 \AA \ (blue band) and from 6400 \AA \ to 6800 \AA \ (red band) were employed for MRS spectra.
The second way involves determining the central wavelength of the $H_{\alpha}$ line by performing a Gaussian fit on the spectral data within the wavelength range of 6555 \AA \ to 6575 \AA.
Then, the RV values can be calculated using the Doppler shift formula with this central wavelength.
Table \ref{rvdata.tab} presents the RV values obtained from these ways, listed in columns 2, 4, and 6.

RV measurements exhibit small systematic offsets due to the temporal variations of zero-points \citep{2019RAA....19...75L,2020ApJS..251...15Z}.
Therefore, a correction corresponding to the zero-point is necessary in order to derive accurate RV values.
In this work, Gaia DR3 data were utilized to determine the RV zero points (RVZPs) for each spectrograph night by night.
The RVZPs were applied as a common RV shift for each fiber in the same spectrograph.
First, we used some criteria ({\it ruwe} $<$ 1.4; {\it radial\_velocity} !$=$ 0; {\it rv\_amplitude\_robust} $<$ 10 km/s; {\it radial\_velocity\_error} $<$ 10 km/s) to exclude possible binaries in the common sources for each spectrograph. 
Then, the same ways were used to calculate the RVs of these sources.
Comparing these RVs with those from Gaia DR3, we estimated the RVZPs and the median offset $\Delta RV$ with 2 or 3 iterations \citep{2021RAA....21..292W}.
In addition, we applied the bootstrap re-sampling method ($N=10000$) to measure the measurement error of RVs.
Finally, we took the square root of the sum of the squares of the measurement error and the median offset $\Delta RV$ as the RV’s uncertainty.
The RV$_{\rm b,cor}$, RV$_{\rm r,cor}$, and RV$_{\rm H_{\alpha},cor}$ in the Table \ref{rvdata.tab} represent the RVs after RVZP correction.

\subsection{Photometry and period measurement} 
\label{sec:phot}

\begin{figure}
    \center
    \includegraphics[width=0.48\textwidth]{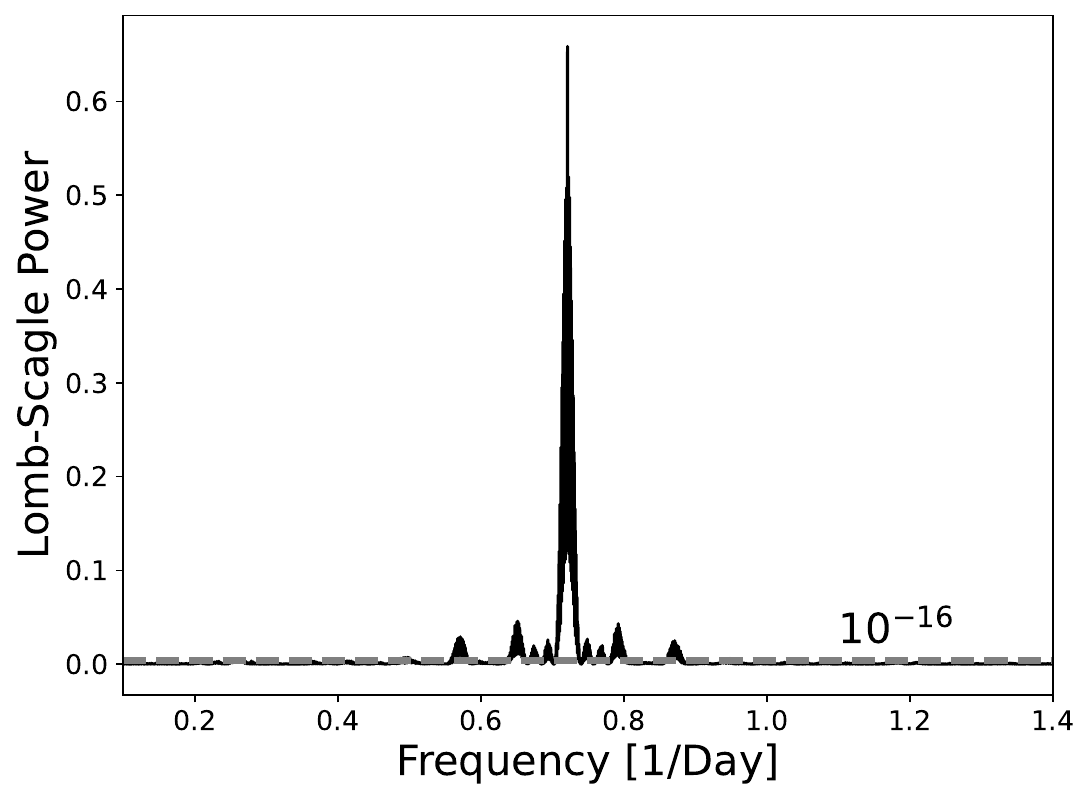}
    \includegraphics[width=0.48\textwidth]{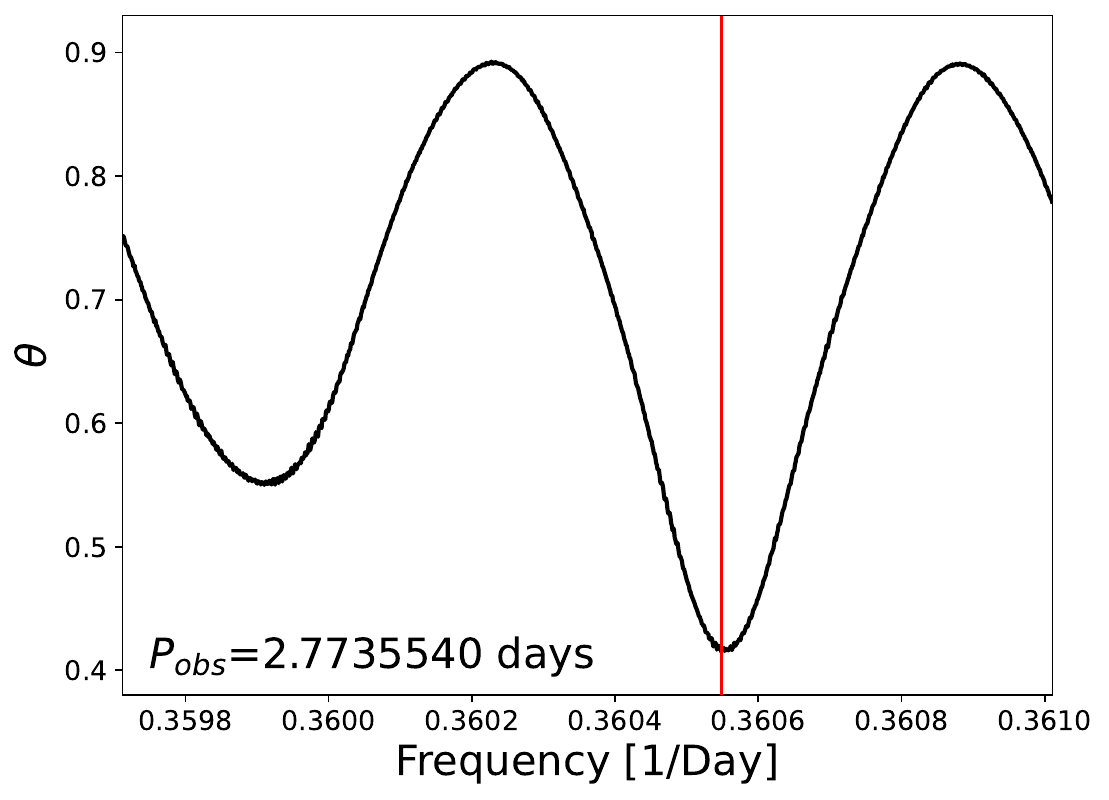}
    \caption{Period estimation of J0606+2132. Top panel: Power spectrum of Lomb-Scargle obtained from the TESS and ASAS-SN $V$- and $g$-band LCs. Bottom panel: The phase dispersion calculated with the TESS and ASAS-SN $V$- and $g$-band LCs. The red line marked the phase dispersion minimum, corresponding to a period of 2.7735540 days.}
    \label{period.fig}
\end{figure}

J0606+2132 has been observed by the All-Sky Automated Survey for Supernovae (ASAS-SN), the Zwicky Transient Facility (ZTF), and the Transiting Exoplanet Survey Satellite (TESS).
We estimated the orbital period using the LCs with two steps (Figure \ref{period.fig}):
a rough estimate using the Lomb-Scargle (LS) method \citep{1989ApJ...338..277P}, followed by a more accurate estimation using the phase dispersion minimization (PDM) method \citep{1978ApJ...224..953S}.
The LCs of TESS and ASAS-SN ($V$ and $g$ bands) were utilized due to their small scatter.
A significant peak around $f=0.721062$ days$^{-1}$ was found in the result of LS, with a probability $<10^{-16}$ that it is due to random fluctuations of photon counts.
In the case of PDM, a search of periods in the frequency range of 0.359971- 0.361011 days$^{-1}$ with a step of $10^{-7}$ days$^{-1}$ returns a phase dispersion minimum at $f=0.3605482$ days$^{-1}$, corresponding to a period of 2.7735540 days.

\begin{figure}
    \center
    \includegraphics[width=0.45\textwidth]{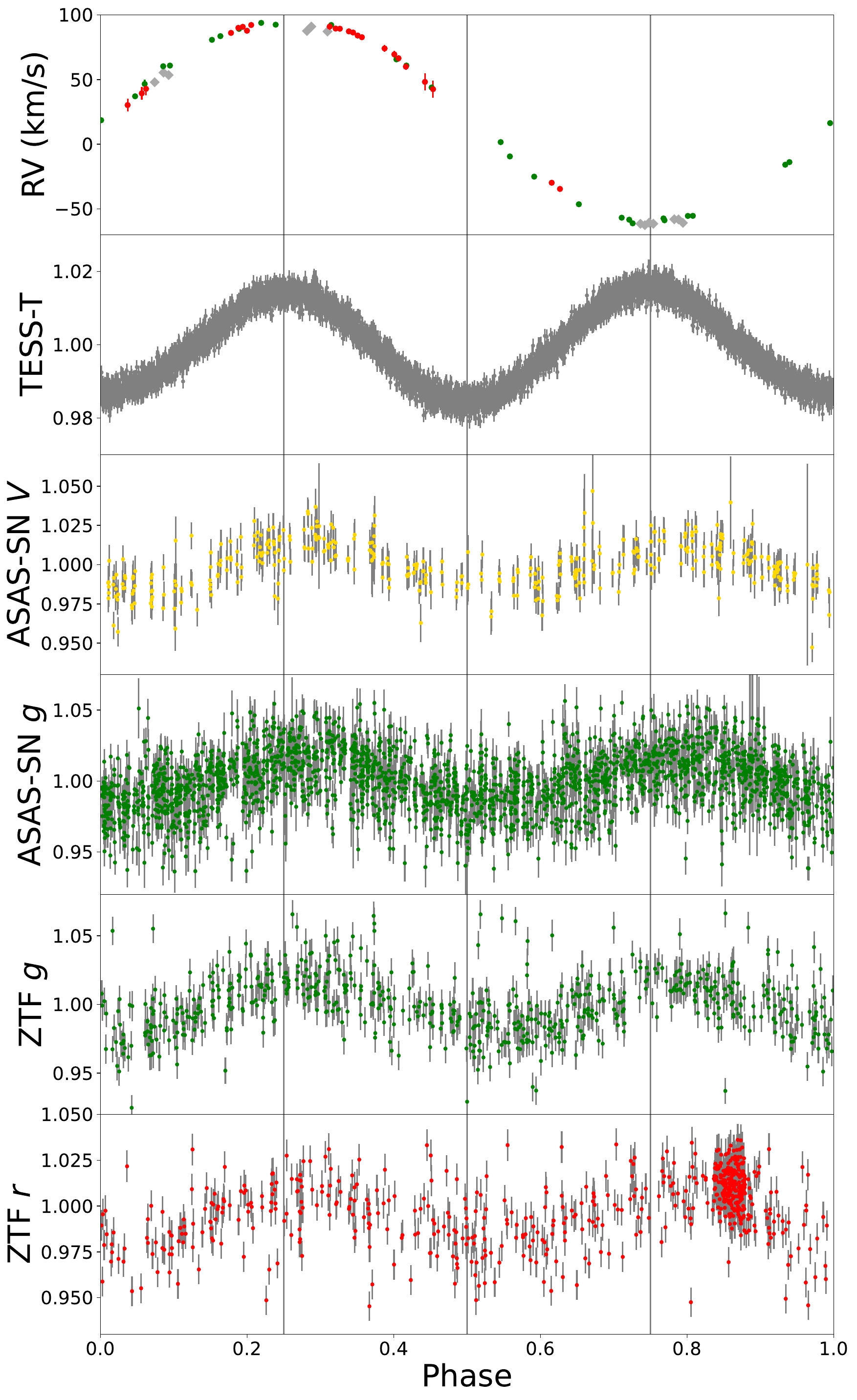}
    \caption{The folded RV data and the normalized LCs in different bands using the period of 2.7735540 days. These vertical gray lines marked the phases of the inferior conjunction, quadrature, and the superior conjunction. The RV data are calculated using the $H_{\alpha}$ line from LAMOST LRS (green dots) and MRS (red dots for S/N$_{r}>$50; grey diamonds for S/N$_{r}<$50) spectra.}
    \label{rv_lcs.fig}
\end{figure}

Figure \ref{rv_lcs.fig} shows the folded RVs and the normalized LCs using the period of 2.7735540 days.
Clear ellipsoidal modulation can be seen in all LCs, indicating one or both stars in the binary are deformed by the tidal forces from their companions.

\section{The Visible Star}
\label{sec:visible_star}

\begin{figure*}
    \center
    \includegraphics[width=0.90\textwidth]{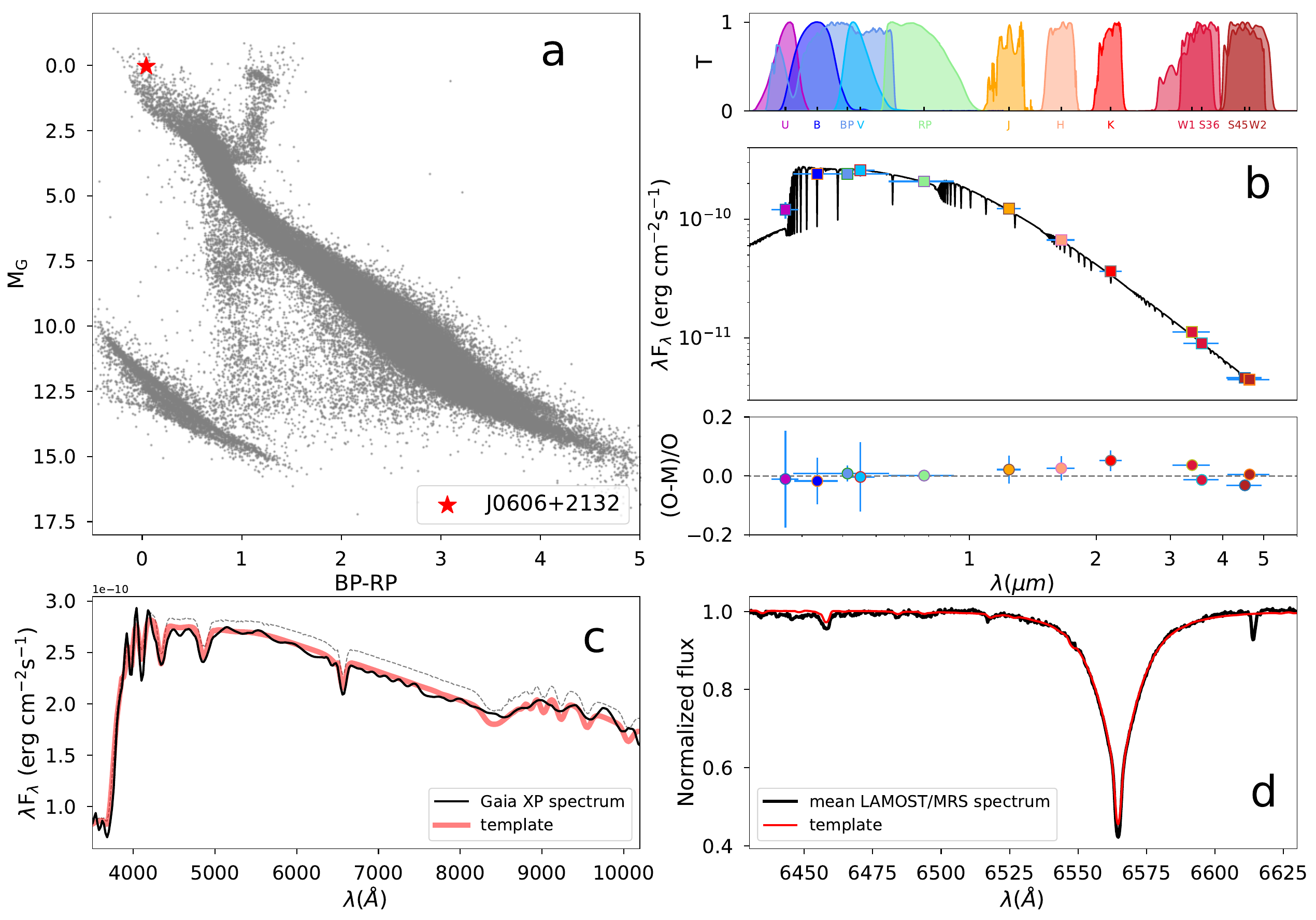}
    \caption{Panel a: Position of the visible star on the Hertzsprung–Russell diagram for J0606+2132. The gray points are plotted for a comparison, which are from {\it Gaia} EDR3 with distances $d <$ 100 pc, $G_{\rm mag}$ between 4--18 mag, and galactic latitudes $|b|$ $>$ 20. No extinction correction was done for these stars. Panel b: SED fitting of J0606+2132. The black line is the best model. Panel c: Comparison of flux-calibrated {\it Gaia} XP spectrum (solid black line) and the Phoenix template ($T_{\rm eff} = 10000$ K, ${\log}g = 3.5$ dex, [Fe/H]$=0$; red line). The flux of the template has been normalized to the distance of J0606+2132, taking extinction into account.
    The dashed line is the combined spectrum of Phoenix templates of the visible star and a main-sequence star with a mass of 1.34 $M_{\odot}$ ($T_{\rm eff} = 6600$ K, ${\log}g = 4.5$ dex, [Fe/H]$=0$). Panel d: Comparison of the mean LAMOST medium-resolution spectrum (black line) and the Phoenix template ($T_{\rm eff} = 10000$ K, ${\log}g = 3.5$ dex, [Fe/H]$=0$, $v{\rm sin}i=67$ km/s; red line; $R\approx 7500$).}
    \label{HR_sed_XP.fig}
\end{figure*}

\subsection{Stellar Parameters}  
\label{sec:params}

The parallax of J0606+2132 is $\varpi=0.5963\pm0.0171$ mas in the Gaia Early Data Release 3 (EDR3) \citep{2021A&A...649A...1G}, corresponding to a distance of $1588.88^{+43.61}_{-39.06}$ pc \citep{2021AJ....161..147B}.
According to the distance, we estimated the extinction of $A_V=1.3$ using a three-dimensional interstellar dust reddening map \citep{2019MNRAS.483.4277C}.
Figure \ref{HR_sed_XP.fig} (panel a) plots its position on the Hertzsprung–Russell diagram, clearly showing that the visible star of J0606+2132 is an early-type star.

The atmospheric parameters ($T_{\rm eff}$, log$g$, and [Fe/H]) of J0606+2132 have been estimated by \cite{2021ApJS..257...22S}.
These authors constructed a data-driven model based on the SLAM model, which was trained with ATLAS12 atmospheric models \citep{2005MSAIS...8..189K}, and derived the atmospheric parameters of $\sim$40000 late-B and A-type main-sequence stars using the LAMOST DR7 MRS data. 
Using the atmospheric parameters derived from multiple observations, we determined the final values for J0606+2132, weighted by the S/N$_r$ of individual spectra.
The atmospheric parameters for J0606+2132 are $T_{\rm eff} = 10233{\pm 382}$ K, log$g$ $= 3.76{\pm 0.21}$, [Fe/H] $= -0.14{\pm 0.11}$, and $v{\rm sin}i=66.88\pm3.59$ km/s (Table \ref{imfos_G3423.tab}), indicating the visible star is a late-B or early-A type star.

It's worth noting that the atmospheric parameters of J0606+2132 were also estimated by \citet{2023ApJS..266...40W}. They reported $T_{\rm eff}$ values from 7200 K to 8000 K and [Fe/H] values between $-$1.1 to $-$0.6 using LAMOST MRS. This discrepancy arises from their use of synthesis spectra of FGKM-type stars ($T_{\rm eff} =$ 3500--8000 K) as training sets, which led to inaccurate parameter estimates for early-type stars.

\begin{table}
\centering
\small
\caption{The physical parameters of J0606+2132.\label{imfos_G3423.tab}}
\setlength{\tabcolsep}{3.5pt}
\begin{tabular}{lll}
\hline\hline
\multicolumn{3}{l}{\bf{Basic properties of J0606+2132}}   \\ 
R.A. & $\alpha$\,[deg] & 91.7051933 \\
Decl. & $\delta$\,[deg] & 21.5425898 \\
Apparent V mag & $V$\,[mag] & 12.22 \\
Extinction & $E(B-V)$\,[mag] & 0.42 \\
\hline
\multicolumn{3}{l}{\bf{Parameters of the visible star}}  \\ 
Effective temperature & $T_{\rm eff,1}$\,[K] & $10233\pm382$ \\
Surface gravity   & $\log(g)_{\rm 1}$ [${\rm cm/s^{2}}$]  & $3.76\pm0.21$  \\
Metallicity & [Fe/H]$_{\rm 1}$ & $-0.14\pm0.11$ \\
Projected RV & $v{\rm sin}i_{\rm 1}$ [km/s] & $66.88\pm3.59$ \\
Radius (SED fitting) & $R_{\rm 1} [R_{\odot}]$ & $3.61^{+0.08}_{-0.08}$  \\ 
Spectroscopic mass &  $M_{\rm 1} [M_{\odot}]$ & $2.69^{+1.67}_{-1.03}$ \\
Evolutionary mass &  $M_{\rm 1} [M_{\odot}]$ & $2.87^{+0.18}_{-0.17}$ \\
\hline
\multicolumn{3}{l}{\bf{RV fitting ({\it The Joker} fitting)}}  \\ 
Period & $P$ [day] & $2.773576^{+0.000021}_{-0.000020}$ \\
Eccentricity   & $e$ [-] & $0.01^{+0.01}_{-0.01}$ \\
Argument of periastron & $\omega$ [radians] & $2.30^{+0.22}_{-0.23}$ \\
Mean anomaly & $M_{\rm 0}$ [radians] & $-0.58^{+0.22}_{-0.23}$ \\
Semi-amplitude of RV & $K$ [km/s] &  $78.03^{+0.22}_{-0.23}$ \\
Systematic RV & $\nu_{\rm 0}$ [km/s] & $18.57^{+0.19}_{-0.19}$ \\
\hline 
\multicolumn{3}{l}{\bf{LC fitting (PHOEBE fitting)}}  \\ 
Effective temperature & $T_{\rm eff,1}$\,[K] & 9950$^{+268}_{-151}$ \\
Mass ratio & $q$ [-] & 2.04$^{+0.38}_{-0.50}$ \\
Inclination angle & $i$ [deg] & 81.31$^{+6.26}_{-7.85}$ \\
Radius of visible star &  $R_{\rm 1} [R_{\odot}]$ & 3.70$^{+0.08}_{-0.05}$ \\
Mass of unseen star &  $M_{\rm 2} [M_{\odot}]$ & 1.34$^{+0.35}_{-0.40}$ \\
\hline
\end{tabular}
\end{table}

\subsection{Spectral Energy Distribution Fitting}
\label{sec:sed}

We performed a spectral energy distribution (SED) fitting using the {\it astroARIADNE} Python module.
The multi-band magnitudes include the $U$ magnitude \citep{2020A&A...638A..18M}, the $B$ and $V$ magnitudes from the fourth United States Naval Observatory CCD Astrograph Catalog (UCAC4), the $BP$ and $RP$ magnitudes from {\it Gaia},  $J$, $H$, and $K_{\rm S}$ magnitudes from the Two Micron All Sky Survey (2MASS), $W$1 and $W$2 magnitudes from the Wide-field Infrared Survey Explorer (WISE), and 3.6 $\mu m$ and 4.5 $\mu m$ magnitudes from Spitzer.
The distance and extinction values were also used as inputs in the SED fitting. 
The atmospheric parameters from SED fitting are
$T_{\rm eff} = 9969^{+118}_{-121}$ K, log$g$ $=$ $3.76^{+0.12}_{-0.12}$, [Fe/H] $= -0.14^{+0.07}_{-0.08}$, and $R_{1}=3.61^{+0.08}_{-0.08} R_{\odot}$, consistent with spectroscopic estimations.
Figure \ref{HR_sed_XP.fig} (panel b) shows the best-fitting in the {\it astroARIADNE}.
There are slight flux excesses for the $J$, $H$, and $K_{\rm S}$ bands, with values of about 2.1\%, 2.5\%, and 5.2\%, respectively.
These excesses are mainly due to the photometric uncertainties in these bands.
Assuming the system is tidally locked, the radius from the SED fitting implies an inclination angle of $i>68^{\circ}$ using the value of $v{\rm sin}i$ ($66.88\pm3.59$ km/s).

\subsection{Mass Determination}
\label{sec:mass}

The spectroscopic mass can be estimated using the stellar radius ($R$) and the surface gravity ($g$) with $M=gR^2/G$, where $R$ was derived from SED fitting and $g$ from spectral analysis. 
We ran 10$^5$ Gaussian sampling iterations of $R$ ($=$3.6$\pm$0.1 $R_{\odot}$) and log$g$ ($=$3.76$\pm$0.21) and calculated the corresponding spectroscopic masses. The 50th, 16th, and 84th percentile of the mass estimates were adopted as the final mass and its lower and upper limits, respectively.
Finally, we determined the spectroscopic mass of J0606+2132 to be $M=2.69^{+1.67}_{-1.03}$ $M_{\odot}$.

In addition to the spectroscopic mass, the mass of the visible star can be estimated as the evolutionary mass by fitting photometric and spectroscopic data to stellar evolutionary models.
We used the {\it isochrones} Python module \citep{2015ascl.soft03010M} to estimate the mass of the visible star, which fits the photometric or spectroscopic parameters with MIST models and returns observed and physical parameters.
The input priors include the effective temperature, surface gravity, metallicity, multi-band magnitudes from $Gaia$ and 2MASS, {\it Gaia} DR3 parallax, and extinction $A_V$. 
The evolutionary mass and radius of J0606+2132 from {\it isochrones} are $M=2.87^{+0.18}_{-0.17} M_{\odot}$ and $R=3.73^{+0.23}_{-0.21} R_{\odot}$, respectively.
The evolutionary mass is consistent with the spectroscopic mass. We used the spectroscopic mass estimates in the following analysis.

\section{Orbital Solution}
\label{sec:orbit}

\subsection{RV Fitting} 
\label{sec:rv_fitting}

\begin{figure*}
    \center
    \includegraphics[width=0.98\textwidth]{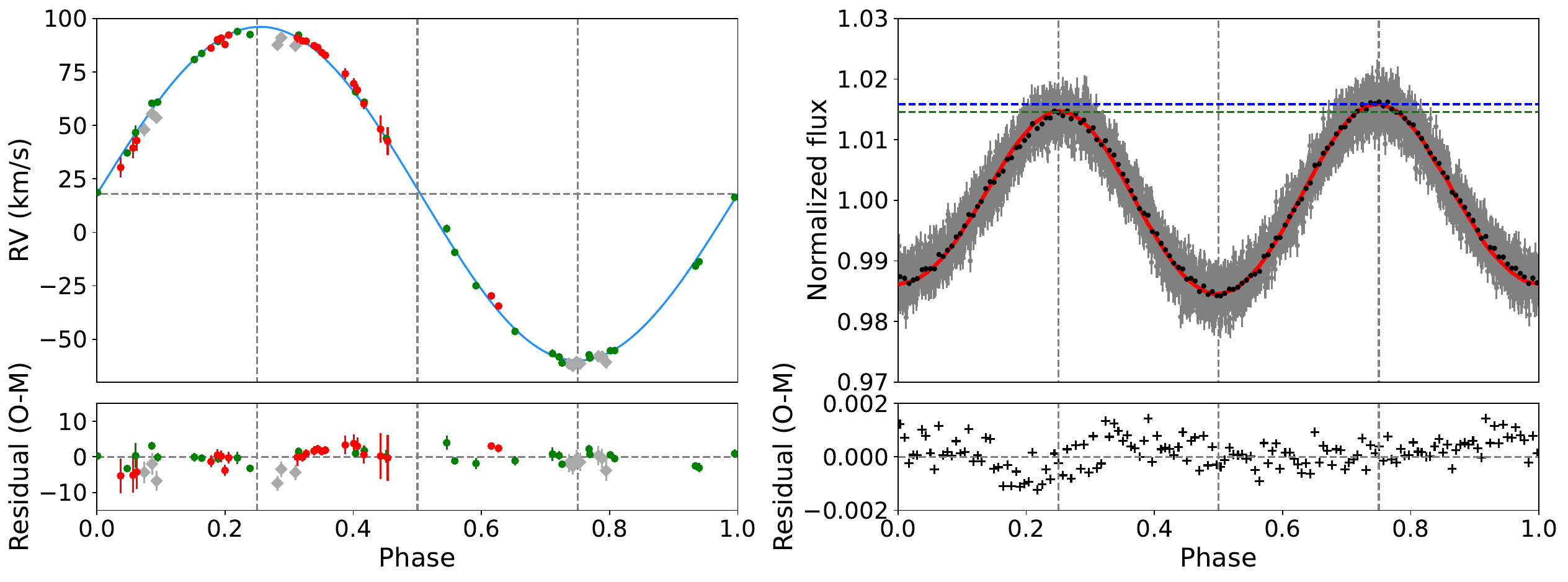}
    \caption{Left panel: Folded RV curve with a period of 2.7735540 days. The phases of the inferior conjunction ($\phi =$ 0 or 1), quadrature ($\phi =$ 0.25 and 0.75), and superior conjunction  $\phi =$ 0.5 are marked with vertical grey lines. The RV data are determined from the LAMOST LRS (green dots) and MRS (red dots for S/N$_{r}>$50; grey diamonds for S/N$_{r}<$50). Right panel: PHOEBE fitting (red line) to the normalized LCs from TESS. The blue dashed line represents the maximum flux at $\phi =$ 0.75, whereas the green dashed line represents the maximum flux at $\phi =$ 0.25. The asymmetric peaks are caused by Doppler boosting.}
    \label{rv_tess.fig}
\end{figure*}

We performed a Keplerian fit using the custom Markov Chain Monte Carlo (MCMC) sampler {\it The Joker} \citep{2017ApJ...837...20P}.
The RV$_{\rm H_{\alpha},cor}$ (Table \ref{rvdata.tab}) with S/N$_{r}>$50 was utilized in the RV fitting due to its lower dispersion compared to RV$_{\rm b,cor}$ and RV$_{\rm r,cor}$.
The fitted orbital parameters from {\it The Joker} are: 
period $P=$2.773576$^{+0.000021}_{-0.000020}$ days,
eccentricity $e = 0.01^{+0.01}_{-0.01}$, 
argument of the periastron $\omega =$2.30$^{+0.22}_{-0.23}$, 
mean anomaly at the first exposure $M_{\rm 0} =$-0.58$^{+0.22}_{-0.23}$, 
semi-amplitude $K =$78.03$^{+0.22}_{-0.23}$ km/s, 
and systematic RV $\nu_{\rm 0} =$18.57$^{+0.19}_{-0.19}$ km/s. 
Figure \ref{rv_tess.fig} shows the folded RV data and the RV curve given by {\it The Joker}.

Consequentially, we calculated the mass function $f(M)$ using these parameters as follows,
\begin{equation}
    f(M) = \frac{M_{2} \, \textrm{sin}^3 i} {(1+q)^{2}} = \frac{P \, K_{1}^{3} \, (1-e^2)^{3/2}}{2\pi G},
\label{eq5}
\end{equation}
\noindent
where $M_{2}$ is the mass of the companion in binary system, $q = M_{1}/M_{2}$ is the mass ratio, and $i$ is the system inclination.
The mass function is $f(M) = 0.136^{+0.001}_{-0.001} M_{\odot}$, corresponding to a companion with the minimum masses ($i=90^{\circ}$) of $1.34^{+0.05}_{-0.05} M_{\odot}$ and $1.29^{+0.42}_{-0.31} M_{\odot}$ for the evolutionary ($2.87^{+0.18}_{-0.17} M_{\odot}$) and spectroscopic (2.69$^{+1.67}_{-1.03} M_{\odot}$) mass of the visible star, respectively.

\subsection{Joint fitting of RV and LC} 
\label{sec:lc_fitting}

Figure \ref{rv_tess.fig} shows that the TESS fluxes vary by about 1.3 parts per thousand ($=\frac{F_2-F_1}{F_2}$) in the phases of the quadrature ($\phi =$ 0.25 and 0.75), which can be attributed to the Doppler beaming effect caused by the orbital motion of the visible star.
\citet{Zheng_2024} presented a straightforward method to calculate the Doppler beaming factor.
They used relativistic theory to calculate the synthetic fluxes by convolving filters in different bands with the spectral template at different RVs.
The Doppler beaming factor was then obtained from a linear fit between these RVs and the ratio between synthetic fluxes and the fluxes at the RV=0 as follows,
\begin{equation}
B = \frac{c}{F_{\rm 0}} \cdot \frac{\partial F_{\rm syn}}{\partial {\rm RV}},
\label{eq:BF}
\end{equation}
where $F_{\rm syn}$ is the synthetic flux at a specific wavelength range, while $F_{\rm 0}$ is the flux at RV=0. $c$ represents the speed of light.
For J0606+2132, the Doppler beaming factor in the TESS-$T$ band is 2.00, calculated using the Phoenix\footnote{ftp://phoenix.astro.physik.uni-goettingen.de/} model with $T_{\rm eff}=10000$ K, log$g$=3.5, and [Fe/H]=0.
The synthetic beaming-flux variation calculated with our RV measurements is about 1.1 parts per thousand, in good agreement with the observed flux variation.

We employed the software Physics of eclipsing binaries \citep{2016ApJS..227...29P, 2018ApJS..237...26H, 2020ApJS..250...34C} (PHOEBE) to generate a grid of LC of J0606+2132 for the following fitting process.
A detached binary configuration was assumed, with a cold ($T_{\rm eff}=300$ K) and small ($R = 3 \times 10^{-6} R_{\odot}$) dark companion.
For the eclipse model of J0606+2132, we used the setting {\it eclipse\_method $=$ only\_horizon}.
For the systematic parameters, the period, eccentricity, and systemic velocity were fixed at $P_{\rm orb}=2.7735540$ days, $e=0$, and $v_{0}=18.57$ km/s, respectively, based on the results from LC and RV analysis.
For the visible star, we fixed the mass at $M_{1}=$2.69 $M_{\odot}$ and adopted Gaussian priors on the temperature ($T_{\rm eff}=9969\pm200$ K), and the radius ($R_{1}=3.6\pm0.2 R_{\odot}$) according to the SED fitting.
The gravity darkening coefficient was set to $\beta=$1\footnote{\url{https://phoebe-project.org/static/legacy/docs/phoebe\_science.pdf}} for stars with radiative envelopes \citep{1924MNRAS..84..665V,1999ASPC..173..277C}.
Moreover, we adopted uniform priors for the mass ratio ($q$) and inclination ($i$) with ranges of [0, 3] and [60$^{\circ}$, 90$^{\circ}$], respectively.

Due to the large dispersion in the ASAS-SN and ZTF LCs, only the TESS LC and RV data were used in the joint fitting.
Furthermore, we found that among the four TESS observations, two (Sectors 71 and 72) show much larger scatter than the other two (Sectors 43 and 44). Therefore, we only used the two LCs with smaller scatter.
In the fitting, we employed an MCMC sampler with 20 walkers and 5000 iterations.
Table \ref{imfos_G3423.tab} lists the results from the best-fit model.
From the best fitting (Figure \ref{rv_tess.fig}), we obtained $T_{\rm eff}=$9950$^{+268}_{-151}$ K, $q=$2.04$^{+0.38}_{-0.50}$, $i=$81.31$^{\circ}$$^{+6.26^{\circ}}_{-7.85^{\circ}}$, and $R_{1}=$3.70$^{+0.08}_{-0.05}\ R_{\odot}$.
The inclination angle is consistent with the value derived from $v{\rm sin}i$ measurement (Section \ref{sec:visible_star}).
The mass of the companion was estimated to be 1.34$^{+0.35}_{-0.40} M_{\odot}$.

\section{Discussion}
\label{sec:discussion}

\subsection{The Nature of the binary}

\subsubsection{Whether the visible star is a stripped star?}  
\label{sec:strip}

Compared to a normal B-type star with a mass of 2.7 $M_{\odot}$, the visible star of J0606+2132 exhibits a slightly larger radius.
Considering this system is a close binary, we investigated the possibility that the visible star is a stripped star.

First, we calculated the size of the Roche lobe using the standard Roche lobe approximation \citep{1983ApJ...268..368E}:
\begin{equation}
\frac{R_1}{a} =  
    \frac{0.49 q^{2/3}}{0.6q^{2/3} + \ln(1 + q^{1/3})},
\label{eq7} 
\end{equation} 
where $a = (1+q) a_1 = (1+q) P (1-e^2)^{1/2} K_1/(2 \pi {\rm \sin}i)$ is the separation of binary system.
Using these results from PHOEBE fitting ($M_{1}$, $M_{2}$) and SED fitting ($R_{1}$), the filling factor of the visible star is $\approx $62$\%$, suggesting the absence of current mass transfer for J0606+2132.

Second, assuming the visible star is a stripped star, we used the PHOEBE to fit the LCs and derive the parameters of both stars to check if they are consistent with observations.
The effective temperature and the radius of the visible star were set to 9969 K and 3.6 $R_{\odot}$, respectively.
For a visible star with a mass less than 0.8 $M_{\odot}$ (e.g., 0.6 and 0.8 $M_{\odot}$), the LC fittings indicate that the Roche lobe of the visible star is smaller than the radius of 3.6 $R_{\odot}$. 
For a visible star with a mass larger than 0.8 $M_{\odot}$ (e.g., 1, 1.2, 1.4, 1.6, and 1.8 $M_{\odot}$), the LC fittings yield a companion with a mass of 3.65, 2.56, 1.98, 1.60, 1.38 $M_{\odot}$ (Table \ref{phoebe_m1.tab}). All of these companions should be detectable from the observed SED or spectra.

Third, we examined the potential presence of a disk using the $H_{\alpha}$ line.
In some binaries including a stripped star \citep{2022MNRAS.516.5945J,2022MNRAS.512.5620E}, their spectra exhibit clear $H_{\alpha}$ emission lines originating from the disk surrounding the accretor.
Figure \ref{Ha.fig} shows the $H_{\alpha}$ profiles of J0606+2132 from MRS spectra with S/N$_{r}>50$, with no emission features at any orbital phase.
However, some binaries undergoing mass transfer exhibit $H_{\alpha}$ absorption lines, such as NGC 1850 BH1 \citep{2022MNRAS.511L..24E}.
This could be due to a very low mass transfer rate or rapid mass loss from the visible star.

\subsubsection{Whether the dark companion is a normal star?}
\label{sec:comp}

We first investigated the possibility that J0606+2132 contains two normal stars.
For the visible star, the absolute $G$- and $RP$-band magnitudes are 0.05 mag and $-$0.07 mag, while the absolute magnitudes of a main-sequence star with a mass of 1.34$M_{\odot}$ are 3.26 mag and 2.93 mag, respectively.
In this case, a flux excess can be detected in the red band of the observed spectrum.
Although there are slight flux excesses in the $J$, $H$, and $K_{\rm S}$ bands in the SED fitting (Figure \ref{HR_sed_XP.fig}, panel b), these can not be caused by a main-sequence companion star, which would contribute $\approx$10\% flux in all three bands. Furthermore, similar flux excesses would also be expected in the WISE and Spitzer bands, but these are not observed in our SED fitting.
Figure \ref{HR_sed_XP.fig} (panel c) shows the comparison of the flux-calibrated Gaia XP spectrum and a Phoenix template with $T_{\rm eff}=10000$ K, log$g=$3.5, and [Fe/H]$=$0.
No clear flux excess is seen, further ruling out the possibility of the dark companion being a late-type main-sequence star.

We applied {\it hoki} \citep{2020JOSS....5.1987S} to search for model binaries including normal stars (i.e., where $M_{2}$ is a visible companion), with observational properties similar to J0606+2132, using the Binary Population and Spectral Synthesis (BPASS) v2.2.1 \citep{2017PASA...34...58E,10.1093/mnras/sty1353}.
Based on the observed metallicity of J0606+2132, we used the binary models with $Z=0.010$.
We performed the selection using the ``primary models" (type$=$1), which consist of normal binaries.
The following search criteria were applied: \\
(i) Orbital period, $P = 2.773\pm0.5$ days; \\
(ii) $J = -0.051\pm0.5$ mag; \\
(iii) $H = -0.068\pm0.5$ mag; \\
(iv) $K = -0.115\pm0.5$ mag; \\
(v) $V = 0.141\pm0.5$ mag; \\
(vi) Effective temperature of the visible star, log($T_{1}$) = $4\pm0.05$; \\ 
(vii) Total luminosity of system, log($L$) = $2.1\pm0.1$; \\
(viii) Radius of the visible star, log($R_{1}$) = $0.56\pm$0.10; \\
(ix) Mass of the visible star, $M_{1}$ = 2--3 $M_{\odot}$; \\
(x) Mass of the companion, $M_{2}$ = 1--2 $M_{\odot}$. \\
No systems with a normal-star companion were found, further suggesting that J0606+2132 is not a normal binary (i.e. stripped star+normal star).

\subsubsection{Spectral disentangling}
\label{sec:sd}

Spectral disentangling is a reliable method to detect faint companions in binaries.
We applied the spectral disentangling algorithm proposed by \cite{1994A&A...281..286S} to the LAMOST MRS spectra.
This algorithm solves for the rest-frame component spectra under the assumption that the component spectra remain constant over time.
If the companion of J0606+2132 is a normal star, the flux contribution in the red band of the spectrum is detectable.
Here we used the wavelength range from 6400 \AA \ to 6600 \AA \ for the spectral disentangling process.

We first did some tests to investigate the detection limit of the companion for J0606+2132.
Assuming the companion is a main-sequence star with different masses ($M=$1.50, 1.35, 1.22$M_{\odot}$ corresponding to mass ratios of $q =$ 1.8, 2.0, 2.2), we derived its atmospheric parameters (i.e., $T_{\rm eff}$ and log$g$)\footnote{\url{http://www.pas.rochester.edu/~emamajek/EEM\_dwarf\_UBVIJHK\_colors\_Teff.txt}}.
The PHOENIX model with the closest atmospheric parameters was used as the model spectrum for the companion.
We re-sampled the model spectra with the resolution of $R=7500$ and applied different rotational broadenings ($v{\rm sin}i$=10 km/s, 50 km/s, 100 km/s, and 150 km/s).
We then generated synthetic binary spectra by combining the observed MRS spectra with S/N$_{r}>$50 and the model spectra following \citep{2022MNRAS.517..356K}:
\begin{equation}
    {f}_{\lambda,{\rm binary}}=\frac{{f}_{\lambda,2} + k_\lambda {f}_{\lambda,1}}{1+k_\lambda},
	\label{eq6}
\end{equation}
where $k_\lambda= \frac{B_\lambda(T{_{\rm eff,1}})~M_1}{B_\lambda(T{_{\rm eff,2}})~M_2} 10^{{\rm log}g_2-{\rm log}g_1}$ is the luminosity ratio per wavelength unit and $B_\lambda$ is the Plank function.
Given that the 6400 \AA \ to 6600 \AA\ range is close to the effective wavelength of $Gaia$ $G$ band, we used the luminosity ratio of $\approx$19 in $G$ band for the disentangling.
For J0606+2132, our method can successfully disentangle the spectra of each component when the mass ratio is below 2.2 and $v \sin i$ is less than 100 km/s. 
Additionally, it can recover the component spectra with low significance when the mass ratio exceeds 2.0 and $v \sin i$ is larger than 100 km/s.

\begin{figure*}
    \center
    \includegraphics[width=0.98\textwidth]{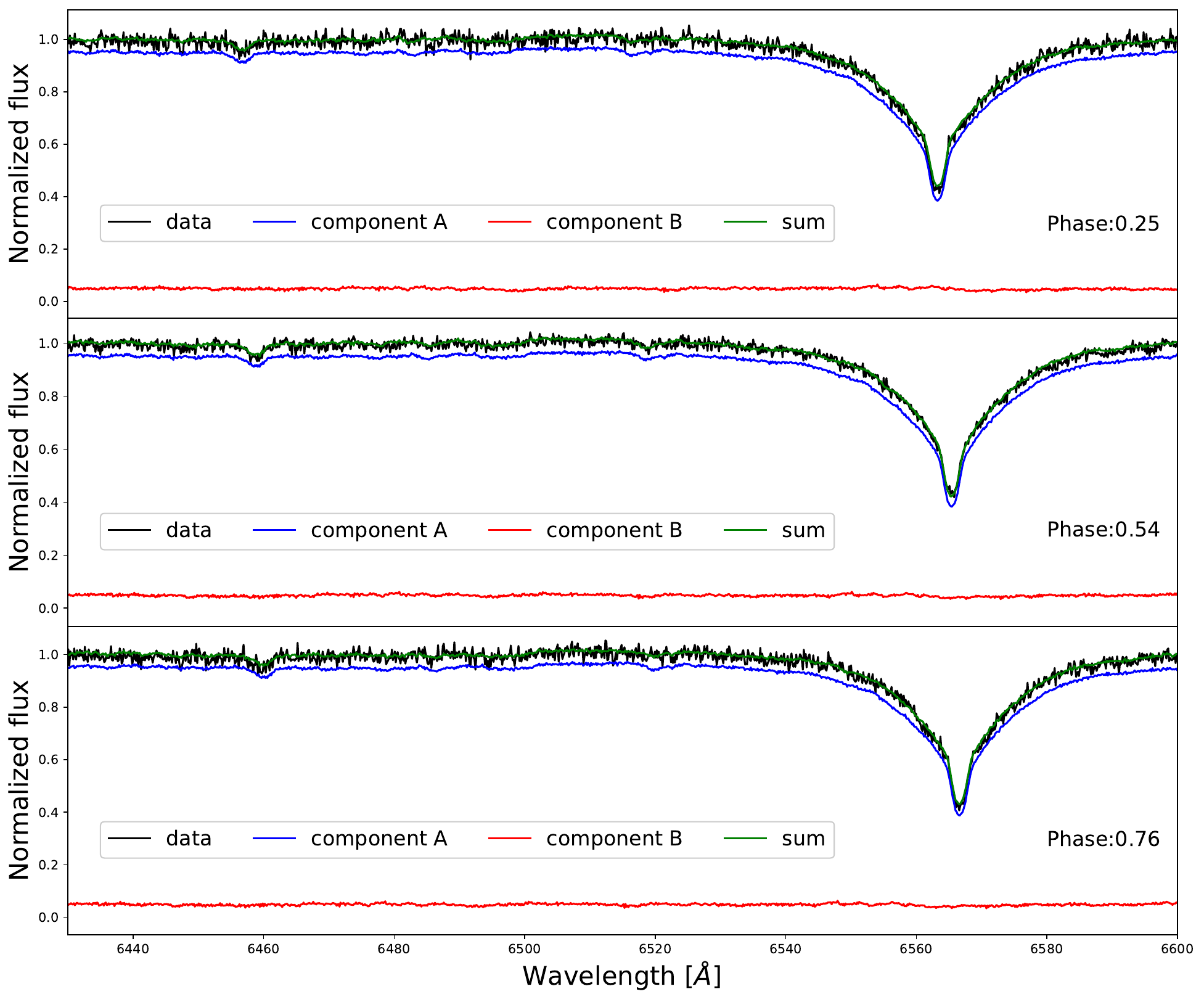}
    \caption{Spectral disentangling with $q =$ 2. The vertical panels show spectra in different phases (close to the minimum or maximum RV phase of the visible star). The blue lines mark the reconstructed spectra of the visible star, while the red lines represent the second component in each spectra. The green lines are the sum of the two components, and the black lines represent the observed spectra.}
    \label{dis_3423.fig}
\end{figure*}

The LAMOST MRS spectra with S/N$_{r}>$50 were used in the final spectral disentangling.
Figure \ref{dis_3423.fig} shows the disentangling results using a mass ratio of $q=2.69/1.34\approx$ 2.
No obvious absorption or emission features can be seen from the spectra of the disentangled component (i.e., component B), suggesting that the companion is a compact object.

\subsection{Evolutionary paths from BPASS}  
\label{sec:bpass}

\begin{figure*}
    \center
    \includegraphics[width=1\textwidth]{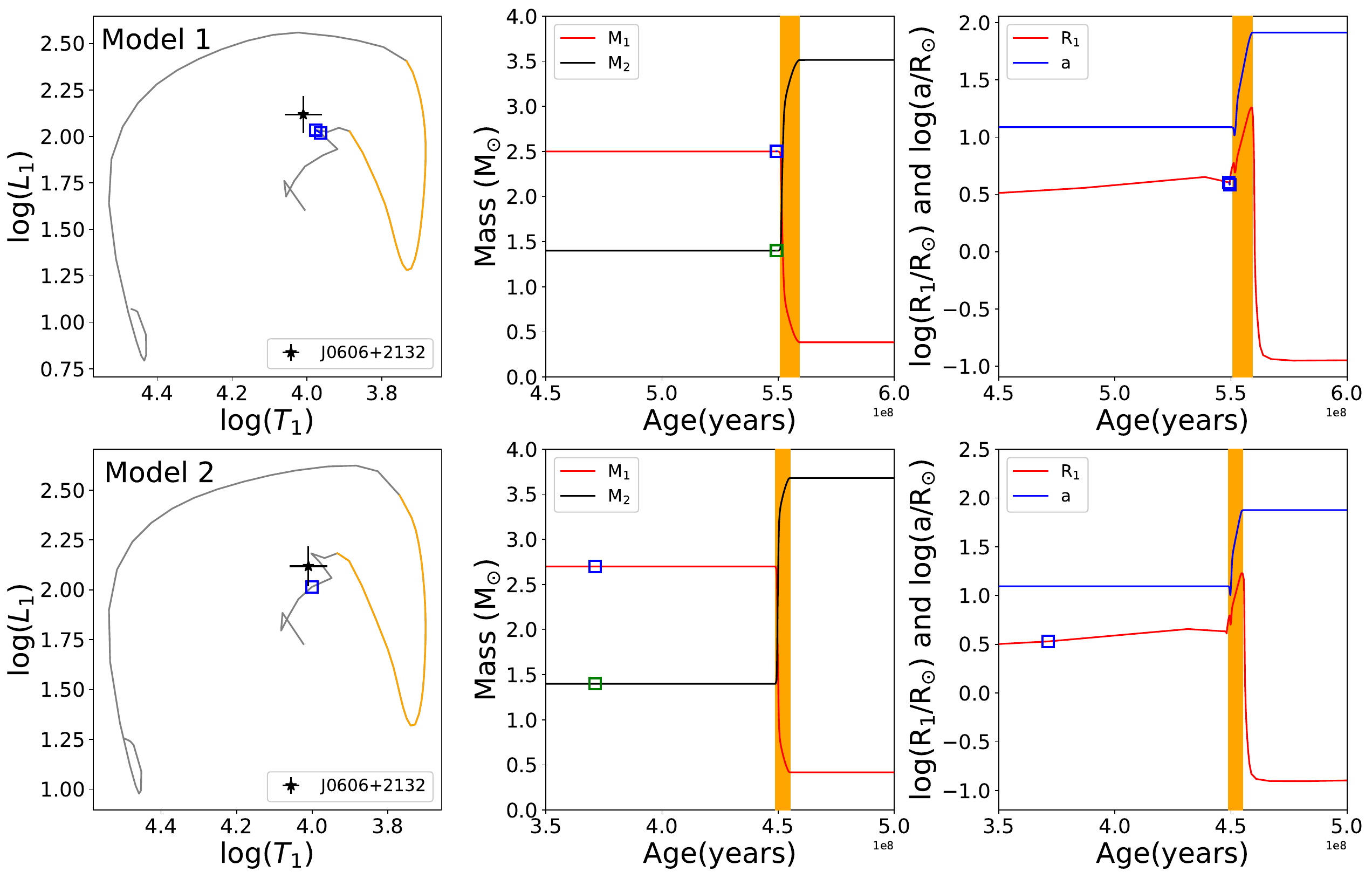}
    \caption{Left panels: Evolutionary tracks of the selected BPASS models in Hertzsprung–Russell Diagrams. The orange lines show the time-steps where RLOF is occurring in these models. The black stars mark the position of J0606+2132. The blue squares are the models matching the search criteria. Middle panels: Evolution of the masses for both components in models. The orange areas represent the evolutionary stages of RLOF. Right panels: Evolution of the binary separation and the radii of the visible stars. The orange areas represent the evolutionary stages of RLOF.}
    \label{bpass_model.fig}
\end{figure*}

To explore the evolutionary paths of J0606+2132, we also searched for binaries that contained compact objects, using the ``secondary models" (type$=$2 in BPASS).
We obtained two models using the same criteria in Section \ref{sec:strip}.
Table \ref{bpass_models.tab} lists basic information for these systems.
Both systems are detached, similar to J0606+2132, suggesting that none of these binaries have yet entered a mass transfer stage.
Figure \ref{bpass_model.fig} shows the evolutionary paths of the two binaries.
If J0606+2132 follows a similar path, its visible star will soon fill its Roche lobe and begin transferring mass to the companion.
Therefore, no X-ray detection of J0606+2132 in current stage is reasonable (Section \ref{sec:Xray_radio}).

Two scenarios are possible for the evolution of J0606+2132.
If the companion is a WD, it may reach the Chandrasekhar limit ($\sim 1.4 M_{\odot}$) by accreting material from the visible star, potentially leading to a Type Ia supernova explosion \citep{2004MNRAS.350.1301H,2023RAA....23h2001L}.
However, if the WD has an ONe core, it may undergo the accretion-induced collapse, forming a NS once enough material is accreted \citep{1991ApJ...367L..19N,2006ApJ...644.1063D,2019ApJ...886..110M}. 
Alternatively, if the companion is an NS, the system could become a candidate for intermediate-mass X-ray binaries (IMXBs), given the mass of the visible star \citep{2000MNRAS.317..438K,2000ApJ...529..946P}.

\begin{table}
\caption{Properties of the nearest matching systems containing compact remnants derived from BPASS models. \label{bpass_models.tab}}
\centering
\setlength{\tabcolsep}{2pt}
\begin{tabular}{cccccc}
\hline\noalign{\smallskip}
Model ID & $M_{1}$ ($M_{\odot}$) & $M_{2}$ ($M_{\odot}$) & $R_{1}$ ($R_{\odot}$) & $T_{1}$ (K) & Age (Myr) \\
\hline\noalign{\smallskip}
1 & 2.5 & 1.4 & 4.03 & 9198 & 549 \\
1 & 2.5 & 1.4 & 3.86 & 9479 & 549 \\
2 & 2.7 & 1.4 & 3.38 & 10020 & 371 \\
\noalign{\smallskip}\hline
\end{tabular}
\end{table}

\subsection{Comparison with other known candidates}

\begin{figure*}
    \center
    \includegraphics[width=0.98\textwidth]{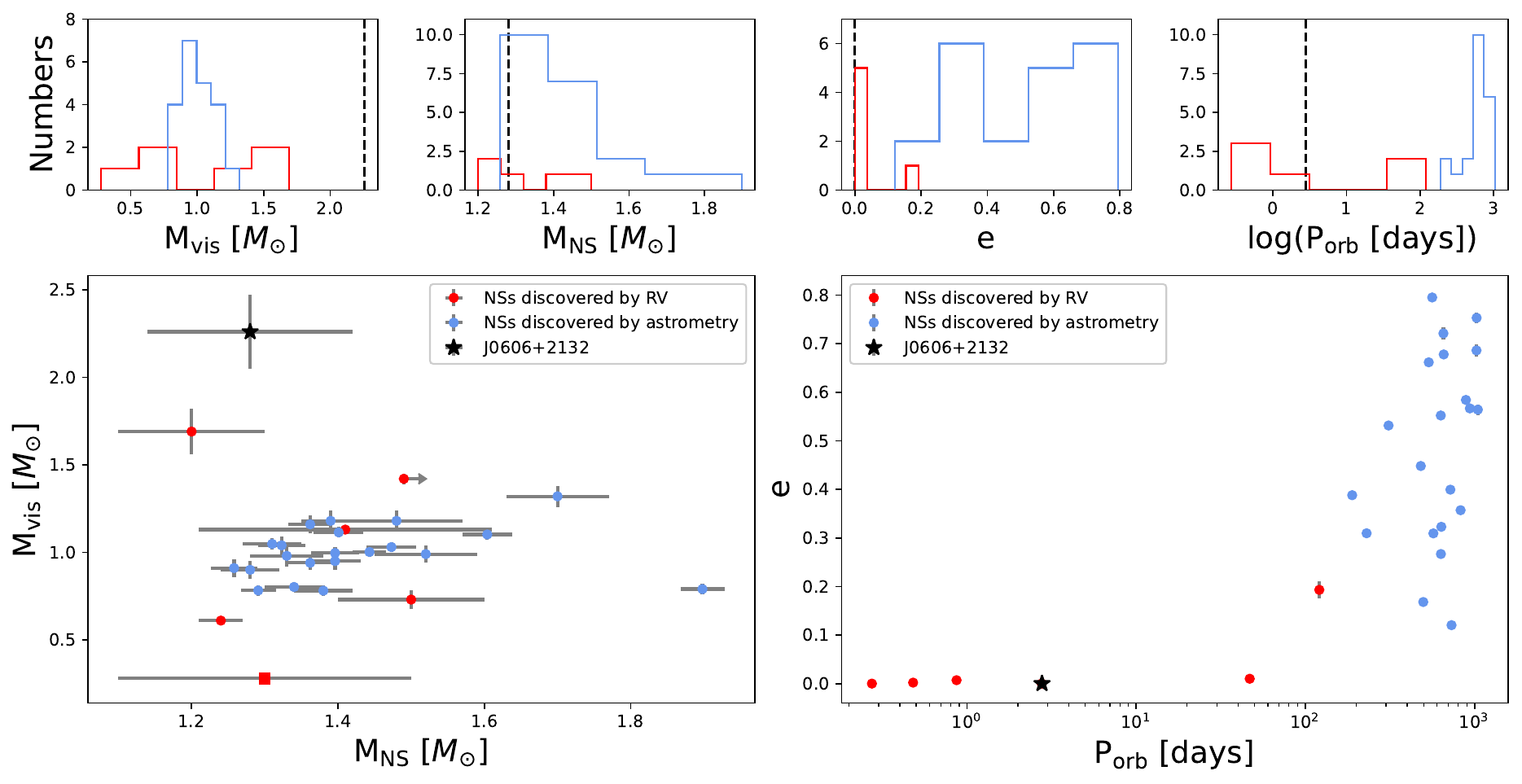}
    \caption{Comparison of J0606+2132 to other NS candidates. Top panels: Histograms of systematic parameters for J0606+2132 and candidates. The blue and red histograms represent systems discovered by astrometry and RV, respectively. Black lines mark the position of J0606+2132 in these histograms. Bottom panels: Masses of visible stars versus masses of the NS candidates (left panel), and eccentricities versus orbital periods (right panel). Red and blue points represent the NSs discovered by RV and astrometry, respectively. The dots and squares represent systems where the visible stars are main sequence stars and giants, respectively.}
    \label{compare_others.fig}
\end{figure*}

We collected other binaries including NS candidates, discovered by RV \citep{2022MNRAS.517.4005M,2022ApJ...940..165Y,2022NatAs...6.1203Y,2023SCPMA..6629512Z,2024ApJ...964..101Z} and astrometry \citep{2024OJAp....7E..58E}, for a comparison.
Figure \ref{compare_others.fig} shows the relations between the masses of the NS candidates and their visible companions, and between the orbital periods and eccentricities.
According to their eccentricities, these systems can be divided into two groups:  close binaries with P$_{\rm orb}\lesssim$ 50 days and circular orbits and the wide binaries with P$_{\rm orb}\gtrsim$ 50 days and eccentric orbits.
The Keplerian orbit of J0606+2132 is similar to that of close binaries, which may have undergone a common envelope phase.
On the other hand, J0606+2132 is unique among these systems since its visible star is an early-type star with a mass greater than 2 $M_{\odot}$, making it a possible candidate for an IMXB.

\subsection{X-ray upper limit and radio observation}  
\label{sec:Xray_radio}

The X-ray sky survey of Roentgen Satellite (hereafter ROSAT) covered the field of J0606+2132, but without a detection of J0606+2132.
The Position Sensitive Proportional Counter (hereafter PSPC) instrument performed one observation of this field with ObsID rs931316n00 and an exposure time of $\sim$ 16417 s.

We used the archival PSPC data to estimate the X-ray upper limit of J0606+2132.
By using aperture photometry with a large radius (100$^{\prime\prime}$), we derived an upper limit of $\approx$ 0.000012 count s$^{-1}$ in the 0.1--2 keV.
The hydrogen column density $N_{\rm H}$ can be calculated through the linear relation between $N_{\rm H}$ and the reddening: $N_{\rm H}=5.8\times10^{21}$ cm$^{-2}$ $E{\rm (B-V)}$.
Assuming different values of the photon index of a power-law spectrum (e.g., $\tau=$1.5, 2), we obtained unabsorbed fluxes of 3.3$\times 10^{-16}\ \rm erg\, s^{-1}\, cm^{-2}$ and 4.8$\times 10^{-16}\ \rm erg\, s^{-1}\, cm^{-2}$, respectively, using WebPIMMS\footnote{https://heasarc.gsfc.nasa.gov/cgi-bin/Tools/w3pimms/w3pimms.pl}.
Given the distance of 1588.88 pc from Gaia EDR3, we further estiamted upper limit X-ray luminosities of 1.0$\times$10$^{29}$ $\rm erg\, s^{-1}$ ($\tau=1.5$) and 1.4$\times$10$^{29}$ $\rm erg\, s^{-1}$ ($\tau=2$).
Therefore, J0606+2132 likely has no or very weak X-ray emission.

J0606+2132 has a Galactic longitude and latitude of 188.7988$^\circ$ and 0.4177$^\circ$. It is within the sky coverage of the Five-hundred-meter Aperture Spherical Telescope \citep[FAST;][]{2019SCPMA..6259502J} Galactic Plane Pulsar Snapshot survey (GPPS) program \citep{2021RAA....21..107H}. The GPPS covers the visible sky of FAST within a Galactic latitude of $\pm 10^\circ$, takes full advantage of the FAST 19 beam observing system, and makes 300s integration for each pointing. It reaches a sensitivity limit of $\mu$Jy, and serves as the deepest Galactic plane pulsar survey program around the world. But it is a pity that no pulsar is found at the source location.  
The non-detection can be explained by either a narrow radio emission beam or the evolutionary paths of binaries involving pulsars.
The companion of a pulsar could be a main-sequence star in two types of systems \citep[e.g.][]{2023pbse.book.....T}. One is for a young pulsar that has not experienced accretion from its main-sequence companion. This type of system generally has a wide and eccentric orbit, i.e., with $e>0.1$ and $P_{\rm orb}>50$ days, and the main-sequence companion is heavy with $m>0.5M_\odot$. The other is for a millisecond pulsar that has experienced accretion from its companion via Case A Roche lobe overflow (RLOF), like low-mass X-ray binaries, in which the companion star evaporation leads to detachment from the Roche lobe. Systems evolved through this channel generally have compact and circular orbits with $e<<0.001$ and $P_{\rm orb}<1$ day, together with a companion of $0.1M_\odot<m<0.4M_\odot$. 
Neither of these two kinds of systems is consistent with J0606+2132.

\section{Summary}  
\label{sec:summary}

We identified a close binary including a late B-type (or early A-type) star and a massive WD or NS.
The atmospheric parameters of the visible star are $T_{\rm eff}=10233\pm382$ K, log$g=3.76\pm0.21$, [Fe/H]=$-0.14\pm0.11$, and $v{\rm sin}i=66.88\pm3.59$ km/s, suggesting a late B-type star with mass of 2.69$^{+1.67}_{-1.03}$.
The SED fitting returns a radius of 3.61$^{+0.08}_{-0.08} R_{\odot}$, slightly larger than normal late B-type stars, indicating that the visible star is evolving off the main sequence.

An obvious ellipsoidal modulation can be found in ASAS-SN, ZTF, and TESS LCs.
We determined an orbital period of 2.7735540 days using the Lomb-Scargle and PDM methods.
Specially, the LCs show asymmetric peaks due to Doppler boosting effect.
By fitting the RV data and LCs, we determined a circular Keplerian orbit, with an inclination of $i=$81.31$^{\circ}$$^{+6.26^{\circ}}_{-7.85^{\circ}}$, which is consistent with the value derived from $v{\rm sin}i$.
Finally, we derived the mass of the unseen object as 1.34$^{+0.35}_{-0.40} M_{\odot}$.

No double-line feature can be found in all observed spectra of LAMOST LRS and MRS.
The comparison of the flux-calibrated Gaia XP spectrum and the Phoenix template suggests that there is no visible companion.
Most notably, spectral disentangling using LAMOST MRS reveals no additional components exhibiting absorption features.
Furthermore, no system with a visible companion is found in BPASS models.
All these results evidence that J0606+2132 contains a compact object, i.e., a massive WD or an NS.

We used BPASS models to explore the evolutionary paths of J0606+2132 and found that it is expected to undergo mass transfer soon.
If the unseen star is a WD, J0606+2132 is a potential progenitor for a Type Ia supernova if it has a CO core, or a candidate for accretion-induced collapse leading to an NS if it has an ONe core, once the WD accretes enough mass to reach the Chandrasekhar limit.
On the other hand, if the unseen star is an NS, J0606+2132 is a plausible candidate for an IMXB scenario when the visible star fills its Roche lobe and starts stable mass transfer via ROLF.

\begin{acknowledgements}

We thank the anonymous referee for very helpful comments and suggestions that significantly improved paper. We thank Prof. Hongwei Ge for very useful discussion on binary evolution. The Guoshoujing Telescope (the Large Sky Area Multi-Object Fiber Spectroscopic Telescope LAMOST) is a National Major Scientific Project built by the Chinese Academy of Sciences. Funding for the project has been provided by the National Development and Reform Commission. LAMOST is operated and managed by the National Astronomical Observatories, Chinese Academy of Sciences. 
This work uses data obtained through the Telescope Access Program (TAP), which has been funded by the TAP member institutes. 
This work presents results from the European Space Agency (ESA) space mission {\it Gaia}. {\it Gaia} data are being processed by the {\it Gaia} Data Processing and Analysis Consortium (DPAC). Funding for the DPAC is provided by national institutions, in particular the institutions participating in the {\it Gaia} MultiLateral Agreement (MLA). The {\it Gaia} mission website is https://www.cosmos.esa.int/gaia. The {\it Gaia} archive website is https://archives.esac.esa.int/gaia. We acknowledge use of the VizieR catalog access tool, operated at CDS, Strasbourg, France, and of Astropy, a community-developed core Python package for Astronomy (Astropy Collaboration, 2013). This research made use of Photutils (Bradley et al. 2020), an Astropy package for detection and photometry of astronomical sources. This work was supported by the National Key Research and Development Program of China (NKRDPC) under grant Nos. 2023YFA1607901, National Science Foundation of China (NSFC) under grant Nos. 11988101/11933004/11833002/12090042/12103047, International partnership program of the Chinese academy of sciences. Grant No. 178GJHZ2022047GC, and Strategic Priority Program of the Chinese Academy of Sciences under
grant No. XDB41000000. 

\end{acknowledgements}

\bibliography{main.bib}
\bibliographystyle{aasjournal}

\clearpage
\appendix
\renewcommand*\thetable{\Alph{section}.\arabic{table}}
\renewcommand*\thefigure{\Alph{section}\arabic{figure}}

\section{Radial velocity measurements}
\label{rvdata_appendix.sec}

\setcounter{table}{0}

Here we present the RV data of J0606+2132 in Table \ref{rvdata.tab}.

\begin{table*}
\caption{Barycentric-corrected RV values of J0606+2132. \label{rvdata.tab}}
\centering
\setlength{\tabcolsep}{1mm}
\renewcommand\arraystretch{0.9}
\begin{center}
\begin{tabular}{ccccccccc}
\hline\noalign{\smallskip}
BJD & RV$_{\rm b}$ & RV$_{\rm b,cor}$ & RV$_{\rm r}$ & RV$_{\rm r,cor}$ & RV$_{\rm H_{\alpha}}$ & RV$_{\rm H_{\alpha},cor}$ & S/N$_{r}$ & Res. \\
(day) & (km/s) & (km/s) & (km/s) & (km/s) & (km/s) & (km/s) & & \\
\hline\noalign{\smallskip}
2458063.32409 & $26.29\pm1.17$ & $41.20\pm1.59$ & $37.31\pm1.01$ & $47.22\pm1.52$ & $39.48\pm2.61$ & $47.99\pm3.06$ & 25.17 & MRS \\
2458063.35743 & $25.79\pm1.30$ & $40.70\pm1.68$ & $45.33\pm1.15$ & $55.24\pm1.62$ & $46.92\pm2.65$ & $55.44\pm3.09$ & 26.50 & MRS \\
2458063.37687 & $30.80\pm1.18$ & $45.71\pm1.59$ & $45.83\pm0.91$ & $55.74\pm1.46$ & $45.04\pm2.23$ & $53.56\pm2.74$ & 28.00 & MRS \\
2458084.24063 & $-38.81\pm0.58$ & $-32.50\pm0.77$ & $-34.81\pm0.55$ & $-30.56\pm0.72$ & $-32.97\pm0.89$ & $-29.72\pm1.04$ & 76.32 & MRS \\
2458084.27188 & $-38.31\pm0.59$ & $-32.00\pm0.78$ & $-39.32\pm0.56$ & $-35.07\pm0.73$ & $-37.73\pm0.93$ & $-34.49\pm1.08$ & 77.11 & MRS \\
2458086.19696 & $82.89\pm0.64$ & $89.36\pm0.69$ & $84.89\pm0.65$ & $89.36\pm0.70$ & $85.52\pm1.04$ & $89.48\pm1.17$ & 69.58 & MRS \\
2458086.21293 & $74.87\pm0.52$ & $81.35\pm0.59$ & $85.39\pm0.70$ & $89.86\pm0.75$ & $85.40\pm1.10$ & $89.36\pm1.23$ & 65.86 & MRS \\
2458086.24696 & $81.39\pm1.07$ & $87.86\pm1.10$ & $83.89\pm0.65$ & $88.36\pm0.70$ & $83.33\pm1.19$ & $87.30\pm1.31$ & 60.68 & MRS \\
2458086.26294 & $82.39\pm0.59$ & $88.86\pm0.65$ & $81.39\pm0.56$ & $85.86\pm0.61$ & $82.45\pm1.01$ & $86.42\pm1.15$ & 69.90 & MRS \\
2458086.28030 & $76.88\pm0.66$ & $83.35\pm0.71$ & $78.88\pm0.72$ & $83.35\pm0.77$ & $80.13\pm1.03$ & $84.09\pm1.17$ & 68.99 & MRS \\
2458086.29627 & $75.38\pm0.72$ & $81.85\pm0.77$ & $78.38\pm0.63$ & $82.85\pm0.68$ & $78.82\pm1.01$ & $82.79\pm1.15$ & 70.80 & MRS \\
2458088.18385 & $19.28\pm0.42$ & $28.36\pm2.46$ & $27.30\pm0.56$ & $32.27\pm5.37$ & $26.54\pm0.94$ & $30.33\pm4.82$ & 72.97 & MRS \\
2458088.23732 & $32.80\pm0.74$ & $41.88\pm2.53$ & $37.31\pm1.10$ & $42.29\pm5.45$ & $35.61\pm0.92$ & $39.40\pm4.82$ & 75.05 & MRS \\
2458088.25329 & $40.32\pm0.42$ & $49.39\pm2.46$ & $36.31\pm0.52$ & $41.29\pm5.37$ & $39.09\pm0.85$ & $42.88\pm4.80$ & 80.37 & MRS \\
2458122.14460 & $85.39\pm0.54$ & $87.40\pm1.60$ & $91.40\pm0.76$ & $90.73\pm1.67$ & $92.44\pm1.48$ & $87.62\pm2.09$ & 49.18 & MRS \\
2458122.16126 & $100.42\pm2.64$ & $102.42\pm3.04$ & $91.90\pm1.04$ & $91.24\pm1.81$ & $95.84\pm1.60$ & $91.02\pm2.18$ & 42.14 & MRS \\
2458122.22237 & $91.40\pm0.62$ & $93.41\pm1.63$ & $88.40\pm0.79$ & $87.73\pm1.68$ & $92.04\pm1.52$ & $87.23\pm2.12$ & 40.72 & MRS \\
2458114.11762 & $68.36\pm0.68$ & $75.96\pm0.98$ & $71.37\pm0.64$ & $75.42\pm1.81$ & $70.82\pm1.11$ & $74.12\pm2.58$ & 63.38 & MRS \\
2458114.15442 & $59.35\pm1.01$ & $66.95\pm1.23$ & $66.36\pm0.68$ & $70.41\pm1.83$ & $66.24\pm1.15$ & $69.54\pm2.59$ & 62.75 & MRS \\
2458114.17039 & $59.35\pm0.95$ & $66.95\pm1.18$ & $64.36\pm0.62$ & $68.40\pm1.80$ & $63.24\pm0.93$ & $66.54\pm2.50$ & 74.42 & MRS \\
2458114.19817 & $48.83\pm0.56$ & $56.43\pm0.90$ & $57.85\pm0.88$ & $61.89\pm1.91$ & $56.74\pm1.06$ & $60.04\pm2.55$ & 65.11 & MRS \\
2458144.04584 & $85.39\pm0.56$ & $92.72\pm0.68$ & $82.39\pm0.82$ & $87.06\pm1.01$ & $81.90\pm1.30$ & $86.12\pm1.66$ & 55.53 & MRS \\
2458144.07362 & $80.38\pm0.70$ & $87.71\pm0.81$ & $83.39\pm1.10$ & $88.06\pm1.26$ & $85.80\pm1.19$ & $90.02\pm1.58$ & 62.72 & MRS \\
2458144.09029 & $81.89\pm2.00$ & $89.21\pm2.04$ & $85.39\pm0.68$ & $90.06\pm0.91$ & $86.58\pm1.25$ & $90.81\pm1.63$ & 59.58 & MRS \\
2458144.10626 & $83.39\pm1.11$ & $90.72\pm1.18$ & $83.89\pm0.87$ & $88.56\pm1.06$ & $83.59\pm1.09$ & $87.81\pm1.51$ & 67.90 & MRS \\
2458144.12223 & $86.39\pm1.03$ & $93.72\pm1.11$ & $86.89\pm0.92$ & $91.57\pm1.10$ & $87.99\pm1.29$ & $92.22\pm1.66$ & 57.32 & MRS \\
2458153.10015 & $45.83\pm0.39$ & $53.29\pm0.92$ & $46.83\pm0.74$ & $50.44\pm5.36$ & $46.33\pm1.19$ & $48.25\pm6.50$ & 61.37 & MRS \\
2458153.13070 & $34.31\pm0.52$ & $41.77\pm0.99$ & $39.82\pm0.76$ & $43.43\pm5.36$ & $40.70\pm1.21$ & $42.62\pm6.50$ & 58.41 & MRS \\
2458186.02236 & $88.90\pm0.70$ & $95.13\pm1.12$ & $87.90\pm0.82$ & $92.01\pm1.78$ & $88.67\pm1.39$ & $91.01\pm2.52$ & 55.89 & MRS \\
2458558.98203 & $-47.83\pm0.66$ & $-47.78\pm1.44$ & $-57.85\pm0.72$ & $-58.02\pm1.93$ & $-57.16\pm1.69$ & $-57.98\pm2.72$ & 49.89 & MRS \\
2458558.99800 & $-61.35\pm0.99$ & $-61.30\pm1.61$ & $-56.84\pm0.72$ & $-57.01\pm1.93$ & $-57.25\pm1.68$ & $-58.08\pm2.72$ & 44.45 & MRS \\
2458559.01466 & $-40.82\pm0.65$ & $-40.77\pm1.43$ & $-57.35\pm0.70$ & $-57.52\pm1.93$ & $-59.91\pm1.92$ & $-60.74\pm2.87$ & 43.77 & MRS \\
2459191.22436 & $-55.84\pm0.83$ & $-55.82\pm0.94$ & $-60.85\pm0.67$ & $-60.50\pm3.23$ & $-61.34\pm1.62$ & $-61.42\pm2.52$ & 47.63 & MRS \\
2459191.24103 & $-66.36\pm0.57$ & $-66.34\pm0.72$ & $-62.85\pm0.70$ & $-62.50\pm3.24$ & $-62.42\pm1.53$ & $-62.50\pm2.46$ & 49.46 & MRS \\
2459191.25700 & $-60.85\pm0.60$ & $-60.83\pm0.75$ & $-58.35\pm0.86$ & $-58.00\pm3.28$ & $-60.51\pm1.71$ & $-60.59\pm2.58$ & 44.84 & MRS \\
2459191.27298 & $-54.84\pm0.58$ & $-54.82\pm0.73$ & $-63.86\pm0.70$ & $-63.51\pm3.24$ & $-61.34\pm1.61$ & $-61.42\pm2.52$ & 47.65 & MRS \\
2457391.11682 & $-55.34\pm1.08$ & $-52.62\pm1.20$ & $-51.84\pm1.84$ & $-56.95\pm1.99$ & $-53.87\pm1.75$ & $-56.73\pm1.90$ & 140.63 & LRS \\
2457716.24128 & $-7.26\pm3.62$ & $-5.01\pm3.68$ & $-15.28\pm1.33$ & $-21.55\pm1.44$ & $-12.68\pm0.49$ & $-15.78\pm0.91$ & 441.49 & LRS \\
2457719.20096 & $26.79\pm3.68$ & $28.69\pm3.74$ & $21.79\pm0.69$ & $15.63\pm0.89$ & $22.21\pm0.41$ & $18.62\pm0.71$ & 512.79 & LRS \\
2457721.21125 & $-51.34\pm3.58$ & $-48.54\pm3.65$ & $-62.85\pm1.40$ & $-67.12\pm1.50$ & $-60.45\pm0.36$ & $-61.11\pm0.70$ & 620.45 & LRS \\
2457724.19353 & $-48.33\pm2.82$ & $-47.20\pm2.90$ & $-55.34\pm1.43$ & $-61.02\pm1.52$ & $-53.29\pm0.34$ & $-55.37\pm0.70$ & 636.70 & LRS \\
2457725.19900 & $89.90\pm2.74$ & $91.42\pm2.82$ & $85.89\pm1.41$ & $81.22\pm1.53$ & $86.12\pm0.33$ & $83.63\pm0.80$ & 642.99 & LRS \\
2457728.18169 & $95.41\pm3.64$ & $99.30\pm3.72$ & $91.40\pm1.16$ & $87.20\pm1.30$ & $93.79\pm0.38$ & $92.48\pm0.80$ & 593.28 & LRS \\
2457740.16176 & $-3.76\pm3.76$ & $-1.04\pm3.91$ & $-8.76\pm1.18$ & $-14.65\pm1.43$ & $-6.66\pm0.43$ & $-9.38\pm0.91$ & 472.34 & LRS \\
2457749.17446 & $-52.34\pm3.82$ & $-49.23\pm3.88$ & $-55.84\pm1.28$ & $-61.44\pm1.53$ & $-54.28\pm0.70$ & $-55.29\pm1.05$ & 277.19 & LRS \\
2457758.15919 & $41.32\pm4.28$ & $43.22\pm4.35$ & $40.82\pm1.18$ & $32.91\pm1.28$ & $42.66\pm0.59$ & $37.11\pm0.98$ & 330.65 & LRS \\
2457759.14744 & $68.86\pm4.18$ & $72.20\pm4.22$ & $71.37\pm1.20$ & $60.92\pm1.31$ & $73.17\pm0.54$ & $65.65\pm0.98$ & 371.32 & LRS \\
2457760.16143 & $-51.84\pm3.68$ & $-48.46\pm3.78$ & $-55.84\pm1.20$ & $-60.69\pm1.40$ & $-55.84\pm0.54$ & $-58.75\pm0.92$ & 367.53 & LRS \\
2457789.06129 & $89.90\pm2.77$ & $94.54\pm3.01$ & $92.40\pm1.25$ & $84.04\pm1.44$ & $94.51\pm0.74$ & $89.22\pm1.14$ & 307.26 & LRS \\
\noalign{\smallskip}\hline
\end{tabular}
\end{center}
\end{table*}

\begin{table*}
\caption{Table \ref{rvdata.tab} (continued).}
\centering
\setlength{\tabcolsep}{1mm}
\renewcommand\arraystretch{0.9}
\begin{center}
\begin{tabular}{ccccccccc}
\hline\noalign{\smallskip}
BJD & RV$_{\rm b}$ & RV$_{\rm b,cor}$ & RV$_{\rm r}$ & RV$_{\rm r,cor}$ & RV$_{\rm H_{\alpha}}$ & RV$_{\rm H_{\alpha},cor}$ & S/N$_{r}$ & Res. \\
(day) & (km/s) & (km/s) & (km/s) & (km/s) & (km/s) & (km/s) & & \\
\hline\noalign{\smallskip}
2457798.01502 & $66.36\pm3.38$ & $71.87\pm3.43$ & $67.36\pm1.00$ & $58.32\pm1.19$ & $68.31\pm0.91$ & $60.90\pm1.34$ & 230.31 & LRS \\
2457798.98748 & $-54.34\pm3.80$ & $-50.83\pm3.99$ & $-51.34\pm1.13$ & $-60.48\pm1.45$ & $-51.31\pm0.69$ & $-57.29\pm1.12$ & 258.29 & LRS \\
2457810.96286 & $65.36\pm3.58$ & $67.49\pm3.70$ & $62.35\pm1.08$ & $57.03\pm1.46$ & $62.72\pm0.46$ & $60.34\pm1.16$ & 476.81 & LRS \\
2457821.98708 & $47.33\pm4.13$ & $51.54\pm4.57$ & $45.33\pm0.60$ & $44.16\pm3.12$ & $47.49\pm0.81$ & $46.65\pm3.49$ & 341.56 & LRS \\
2457827.97495 & $90.90\pm3.80$ & $95.13\pm3.97$ & $93.41\pm1.17$ & $89.64\pm1.57$ & $94.43\pm0.74$ & $93.89\pm1.78$ & 247.92 & LRS \\
2457839.97482 & $2.25\pm3.78$ & $6.78\pm4.20$ & $0.25\pm1.12$ & $-2.70\pm1.82$ & $2.76\pm0.66$ & $1.70\pm1.98$ & 322.93 & LRS \\
2458076.21315 & $-53.84\pm3.66$ & $-53.84\pm3.79$ & $-56.84\pm1.33$ & $-63.21\pm1.51$ & $-57.18\pm0.55$ & $-58.28\pm1.17$ & 411.59 & LRS \\
2458077.24960 & $70.37\pm3.98$ & $72.37\pm4.07$ & $63.86\pm1.24$ & $58.81\pm1.41$ & $63.98\pm0.78$ & $60.87\pm1.20$ & 282.38 & LRS \\
2458099.16145 & $21.29\pm3.70$ & $23.64\pm3.90$ & $20.78\pm1.32$ & $13.27\pm1.59$ & $20.21\pm0.55$ & $16.39\pm1.21$ & 427.35 & LRS \\
2458105.14407 & $86.89\pm3.65$ & $89.76\pm3.80$ & $84.89\pm1.24$ & $78.05\pm1.56$ & $84.65\pm0.57$ & $80.78\pm1.28$ & 411.48 & LRS \\
2458109.13634 & $-19.78\pm3.74$ & $-16.80\pm3.86$ & $-23.79\pm0.90$ & $-30.75\pm1.25$ & $-22.12\pm0.53$ & $-24.99\pm1.56$ & 415.89 & LRS \\
2458135.06377 & $-1.25\pm3.80$ & $-0.14\pm3.89$ & $-14.77\pm1.25$ & $-19.64\pm1.59$ & $-13.92\pm0.63$ & $-13.75\pm1.30$ & 359.70 & LRS \\
2458162.00285 & $-45.33\pm4.86$ & $-40.65\pm5.03$ & $-44.82\pm1.10$ & $-50.36\pm1.53$ & $-45.06\pm0.73$ & $-46.34\pm1.28$ & 313.78 & LRS \\
2458788.26919 & $38.81\pm5.32$ & $50.42\pm5.38$ & $34.31\pm1.25$ & $38.01\pm1.43$ & $37.48\pm1.56$ & $43.93\pm1.82$ & 182.58 & LRS \\
2459614.40849 & $90.90\pm0.88$ & $98.80\pm0.99$ & $90.90\pm2.00$ & $92.36\pm2.02$ & $90.92\pm1.17$ & $92.22\pm1.24$ & 240.10 & LRS \\
\noalign{\smallskip}\hline
\end{tabular}
\end{center}
\end{table*}

\section{The observed $H_{\alpha}$ profiles}
\label{Ha_dis.sec}

Figure \ref{Ha.fig} shows the observed $H_{\alpha}$ profiles from LAMOST MRS in different phases. 
At phase $\sim$0 (the two lowest red lines in Figure \ref{Ha.fig}), the $H_{\alpha}$ profiles seem to show an emission feature, or an absorption feature from an additional component (e.g, a normal star).
However, if this feature were indeed from an additional component, it should be clearly visible in the spectra at phases around 0.25, where the binary spectra are more distinctly separated.
Additionally, another spectrum at phase$\sim$0 exhibits a single line feature, further suggesting the features in the other two spectra are caused by noise.

\setcounter{figure}{0}
\setcounter{table}{0}

\begin{figure}
    \center
    \includegraphics[width=0.48\textwidth]{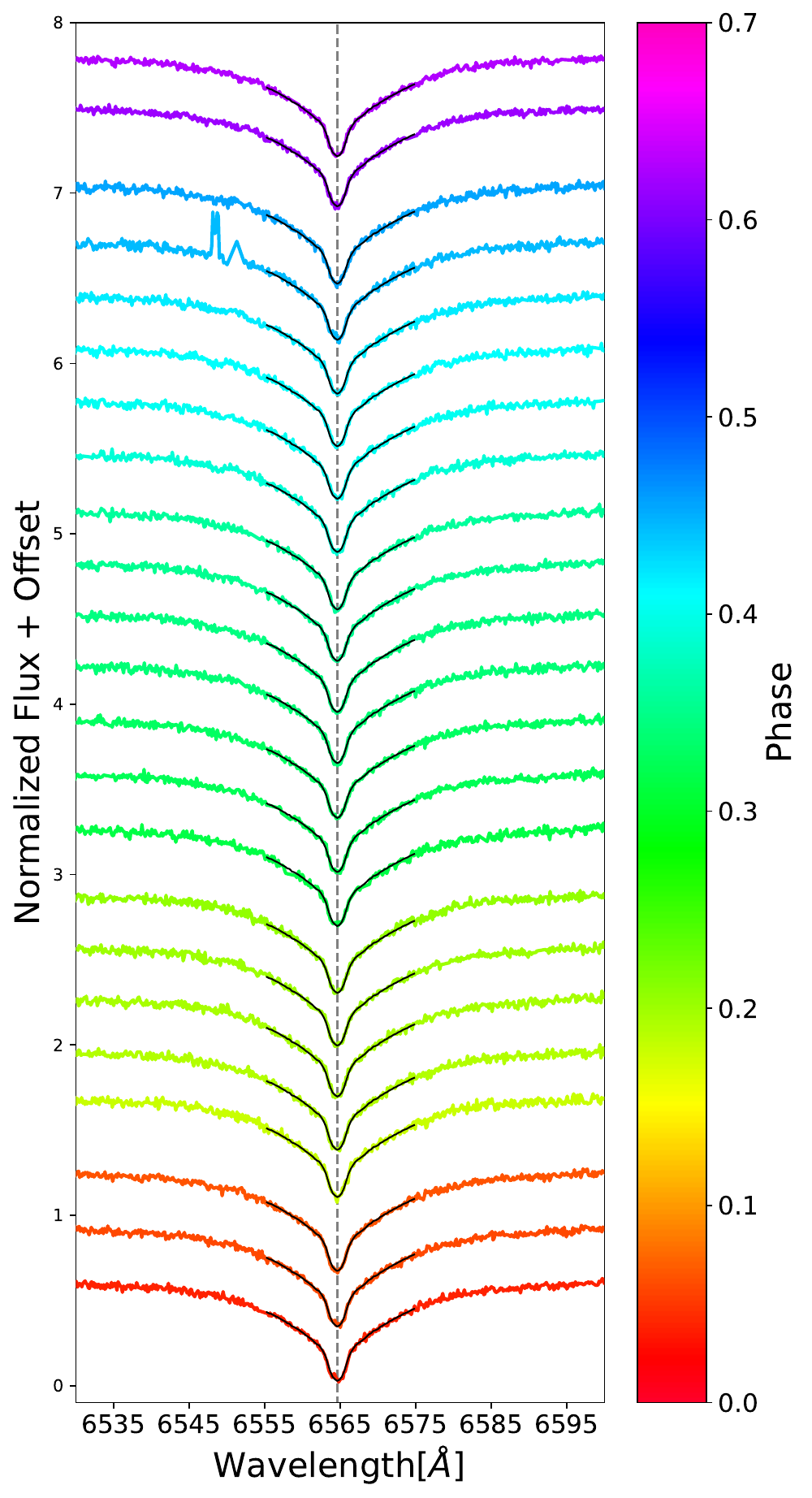}
    \caption{$H_{\alpha}$ line profiles from LAMOST MRS observations plotted in different phases. The black lines show the mean spectrum of those MRS spectra.}
    \label{Ha.fig}
\end{figure}

\section{PHOEBE parameter estimates for different masses of visible stars}
\label{.sec}

\setcounter{table}{0}

Based on the current orbital parameters and the observed radius of the visible star, we estimated the mass of the companion for different masses of visible stars by LC fitting.
The period, eccentricity, and the systematic velocity also were fixed ($P_{\rm orb}=2.7735540$ days, e=0, and $v_{0}=18.57$ km/s), while the radius of the visible star was set at $3.6 R_{\odot}$ according to the SED results.
Table \ref{phoebe_m1.tab} shows the results of PHOEBE parameter estimates for different masses of visible stars.

\begin{table}
\caption{The results of PHOEBE parameter estimates. \label{phoebe_m1.tab}}
\centering
\setlength{\tabcolsep}{2mm}
 \begin{tabular}{cccc}
\hline
$M_{1}$ ($M_{\odot}$) & $i$ ($ \ ^{\circ}$) & $q$ & $M_{2}$ ($M_{\odot}$) \\
\hline
0.6 & - & - & - \\
0.8 & - & - & - \\
1.0 & $23.60^{+0.57}_{-0.52}$ & $0.27^{+0.03}_{-0.03}$ & $3.65^{+0.42}_{-0.37}$ \\
1.2 & $29.29^{+0.89}_{-0.85}$ & $0.47^{+0.06}_{-0.05}$ & $2.56^{+0.31}_{-0.28}$ \\
1.4 & $35.74^{+1.60}_{-1.34}$ & $0.71^{+0.09}_{-0.08}$ & $1.98^{+0.25}_{-0.23}$ \\
1.6 & $43.98^{+2.21}_{-1.82}$ & $1.00^{+0.13}_{-0.09}$ & $1.60^{+0.16}_{-0.18}$ \\
1.8 & $53.65^{+3.71}_{-3.16}$ & $1.31^{+0.16}_{-0.13}$ & $1.38^{+0.16}_{-0.15}$ \\
\hline
\end{tabular}
\end{table}

\end{document}